\documentclass[12pt]{article}

\usepackage{graphicx}
\usepackage{dcolumn}
\usepackage{bm}
\usepackage{amsmath}
\usepackage{epstopdf}
\usepackage{amsfonts}
\usepackage{amssymb}
\usepackage{epsfig}
\usepackage{tabularx}
\usepackage{color}
\usepackage{cite} 
\usepackage{float}
\usepackage{xcolor}

\usepackage[colorlinks=true,citecolor=blue,urlcolor=magenta,breaklinks]{hyperref}

\newcommand{\be}{\begin{equation}}
\newcommand{\en}{\end{equation}}
\newcommand{\bea}{\begin{eqnarray}}
\newcommand{\ena}{\end{eqnarray}}

\begin{document}

\begin{titlepage}



\centerline{\large \bf {Quasinormal modes for a non-minimally coupled scalar field}}

\centerline{\large \bf {in a five-dimensional Einstein-power-Maxwell background}}

\vskip 1.0cm

\centerline{\'Angel Rinc{\'o}n$^{\clubsuit}$, P. A. González$^{\diamondsuit}$, Grigoris Panotopoulos$^{\spadesuit}$,  Joel Saavedra$^{\star}$, Yerko Vásquez$^{\triangle}$, }

\vskip 0.7cm

\centerline{Sede Esmeralda, Universidad de Tarapac\'a,}
\centerline{Avda. Luis Emilio Recabarren 2477, Iquique, Chile.}
\centerline{$^{\clubsuit}$email:
\href{mailto:aerinconr@academicos.uta.cl}
{\nolinkurl{aerinconr@academicos.uta.cl}}
}

\vskip 0.3cm

\centerline{Facultad de Ingenier\'ia y Ciencias, Universidad Diego Portales,}
\centerline{Avenida Ej\'ercito Libertador 441, Casilla 298-V, Santiago, Chile.}
\centerline{$^{\diamondsuit}$email:
\href{mailto:pablo.gonzalez@udp.cl}
{\nolinkurl{pablo.gonzalez@udp.cl}}
}

\vskip 0.3cm

\centerline{Departamento de Ciencias F{\'i}sicas, Universidad de la Frontera, } \centerline{Casilla 54-D, 4811186 Temuco, Chile.}
\centerline{$^{\spadesuit}$email:
\href{mailto:grigorios.panotopoulos@ufrontera.cl}
{\nolinkurl{grigorios.panotopoulos@ufrontera.cl}}
}

\vskip 0.3cm

\centerline{
Instituto de F{\'i}sica, Pontificia Universidad Cat{\'o}lica de Valpara{\'i}so,} 
\centerline{Casilla 4950, Valpara{\'i}so, Chile,}
\centerline{$^{\star}$email:
\href{mailto:joel.saavedra@pucv.cl}
{\nolinkurl{joel.saavedra@pucv.cl}}
}

\vskip 0.3cm

\centerline{
Departamento de F{\'i}sica, Facultad de Ciencias, Universidad de La Serena,} 
\centerline{Avenida Cisternas 1200, La Serena, Chile.}
\centerline{$^{\triangle}$email:
\href{mailto:yvasquez@userena.cl}
{\nolinkurl{yvasquez@userena.cl}}
}

\vskip 1.0cm

\begin{abstract}

We study  the propagation of massless scalar fields non-minimally coupled to gravity  in the background of five-dimensional Einstein-power-Maxwell black holes, and using the WKB and the pseudospectral Chebyshev methods we obtain the quasinormal frequencies (QNFs), which allow us to show the existence of stable and unstable QNFs depending on the power $k$ of the nonlinear electrodynamics, the $\xi$ parameter, which controls the strength of the non-minimal coupling, and on the size of the black hole.
\end{abstract}

\end{titlepage}

\section{Introduction}


The quasinormal modes (QNMs) and quasinormal frequencies (QNFs)
\cite{Regge:1957td,Zerilli:1971wd, Kokkotas:1999bd, Nollert:1999ji, Konoplya:2011qq,Berti:2009kk}
have recently acquired great interest due
to the detection of gravitational waves \cite{Abbott:2016blz}.
Despite the detected signal is consistent with the Einstein gravity \cite{TheLIGOScientific:2016src}, there are possibilities for alternative theories of gravity due to the large uncertainties in mass and angular momenta of the ringing black hole \cite{Konoplya:2016pmh}. In this context, extensive studies of QNMs of black holes in
asymptotically flat spacetimes have been performed for the last few decades mainly due to the potential astrophysical interest.\\

QN spectra of black holes have been investigated in detail from long time ago, although recently, the interest in such a type of research is higher than ever. In particular, the exact  analytical calculations (for quasinormal spectra of black holes) is only possible to get in a concrete number of cases. To name a few:
i) when the effective potential barrier acquire the (simple) form of the P{\"o}schl-Teller potential \cite{potential,ferrari,cardoso2,lemos,molina,panotop1}, 
or 
ii) when the corresponding differential equation (for the radial part of the wave function) can be rewritten into the Gauss' hypergeometric function \cite{exact1,exact2,exact3,Gonzalez:2010vv,exact4,exact5,exact6}. 
In general, an exact solution is not possible to achieve (due to the complexity and non-trivial structure of the differential equation involved), reason why it is necessary to employ some numerical method. 
Up to now we have a variety of methods used to obtain, in a good approximation, the corresponding QNMs of black holes.
Jut to mention a few of them, we have the Frobenius method, generalization of the Frobenius series, fit and interpolation approach, method of continued fraction, among others. 
Additional details can be consulted, for example, in \cite{review3}. \\

On the other hand, although our observable Universe is clearly four-dimensional, the question "How many dimensions are there?" is one of the fundamental questions that modern High Energy Physics tries to answer. Kaluza-Klein theories \cite{kaluza,klein}, Supergravity \cite{nilles} and Superstring/M-Theory \cite{ST1,ST2} have pushed forward the idea that extra spatial dimensions may exist. Here, we consider as background five-dimensional black hole solutions for the power Maxwell theory coupled to gravity \cite{Hendi:2012um, Zangeneh:2015wia}, and we study the propagation  of massless scalar fields, by using, the Wentzel-Kramers-Brillouin (WKB) approximation  \cite{wkb1,wkb2,wkb3}, which have been successful applied to different circumstances. An partial and incomplete list is, for instance: \cite{paper1,paper2,paper3,paper4,paper5,paper6}, and for more recent works \cite{paper7,paper8,paper9,paper10,Rincon:2018sgd,Panotopoulos:2020mii,Rincon:2020cos,Rincon:2020pne,Rincon:2020iwy,Panotopoulos:2019gtn,Panotopoulos:2019qjk}, and references therein. Also, we use the pseudospectral Chebyshev method \cite{Boyd}, which is an effective method to find high overtone modes and  have been successful applied to some spacetimes \cite{Finazzo:2016psx,Gonzalez:2017shu,Gonzalez:2018xrq,Becar:2019hwk,Aragon:2020qdc, Aragon:2020tvq, Aragon:2020xtm, Aragon:2020teq, Fontana:2020syy}. \\

The advantage of the Einstein-power-Maxwell (EpM) theory is that it preserves the nice conformal properties of the four-dimensional Maxwell's theory in any number of spacetime dimensionality $D$.
Regular black hole solutions have been reported in nonlinear electrodynamics \cite{Bardeen, AyonBeato:1998ub, AyonBeato:1999rg, Cataldo:2000ns, Bronnikov:2000vy, Burinskii:2002pz, Matyjasek:2004gh}. Besides,  higher dimensional black hole solutions to Einstein-dilaton theory coupled to the Maxwell field were found in \cite{Hendi:2015xya, Dehghani:2006zi} and black hole solutions to Einstein-dilaton theory coupled to Born-Infeld and power-law electrodynamics were found in \cite{Zangeneh:2015wia}.\\

Our work in the present article is organized as follows: In the next section we briefly review charged BH solutions in EpM theory, and we also very briefly discuss the wave equation with the corresponding effective potential well and  potential barrier for the scalar perturbations. In the third section we compute the QNFs adopting the WKB approximation of 6th order, for stable modes, and the pseudospectral Chebyshev method for unstable modes, and we discuss our results. Finally, in section four we summarize our work with some concluding remarks. We adopt the mostly positive metric signature $(-,+,+,+,+)$, and we work in geometrical units where the universal constants are set to unity, $c=1=G_5$.

\section{Background and scalar perturbations}

\subsection{Charged black hole solutions in EpM theory}

We will investigate a 5-dimensional theory parameterized by the action
\begin{equation}
S[g_{\mu \nu}, A_\mu] = \int \mathrm{d} ^5x \sqrt{-g} \left[ \frac{1}{2 \kappa} R - \alpha | F_{\mu \nu} F^{\mu \nu} |^k \right]\,,
\end{equation}
where $R$ is the Ricci scalar, $g$ the determinant of the metric tensor $g_{\mu \nu}$, $A_\mu$ is the Maxwell potential,  $\kappa=8 \pi$, $k$ is an arbitrary rational number and $F \equiv F_{\mu \nu} F^{\mu \nu}$ is the Maxwell invariant with $F_{\mu \nu}$ being the electromagnetic field strength, which is defined as
\begin{equation}
F_{\mu \nu} \equiv \partial_\mu A_\nu - \partial_\nu A_\mu\,,
\end{equation}
where the indices run from 0 to $4$.\\

It is possible to consider the absolute value in the action for the Maxwell invariant (i.e., $\mathcal{L}_{EM} = -\alpha|F|^k, \forall	\ \alpha \in \mathbb{R} $), which ensures that any configuration of electric and magnetic fields can be described by this Lagrangian, or alternatively, one could consider the Lagrangian without the absolute value and the exponent $k$ restricted to being an integer or a rational number with an odd denominator \cite{EpM1}, in this case the Lagrangian density becomes of the form $\mathcal{L}(F) \sim F^k$.
To obtain the corresponding equations of motion, we first vary the action with respect to the metric tensor sourced by the electromagnetic energy-momentum tensor \cite{EpM1}
\begin{equation}
G_{\mu \nu}  =  4 \kappa \alpha \left [k F_{\mu \rho} F_\nu ^\rho F^{k-1} - \frac{1}{4} g_{\mu \nu} F^k \right ]\,,
\end{equation}
where $G_{\mu \nu}$, as always, is the Einstein tensor. Second, varying the action with respect to the Maxwell potential $A_\mu$ we obtain the generalized Maxwell equations \cite{EpM1,EpM4,EpM10,Rincon:2021hjj}, namely:
\begin{equation}
\partial_\mu (\sqrt{-g} F^{\mu \nu} F^{k-1}) = 0\,.
\end{equation}
We start by considering a static, spherically symmetric solution taking  the metric tensor as
\begin{equation}
ds^2 = -f(r) dt^2 + f(r)^{-1} dr^2 + r^2 d \Omega_{D-2}^2\,,
\end{equation}
where $r$ is, as always, the radial coordinate, and with $d \Omega_{3}^2$ is  the line element of the unit 3-dimensional sphere\cite{EGB1,EGB2}
\begin{equation}
\mathrm{d} \Omega_3^2 = \mathrm{d} \theta^2 + \sin^2 \theta \mathrm{d} \phi^2 + \sin^2 \theta \sin^2 \phi \mathrm{d}\psi^2\,.
\end{equation}
Also, the electric field $E(r)$ is found to be \cite{EpM1}
\begin{equation}
E(r)=F_{rt} = \frac{C}{r^\beta}\,.
\end{equation}
Here we have two constants: i)  $C$ is a constant of integration, and ii) is the exponent $\beta$, which is given by \cite{EpM1}
\begin{equation}
\beta = 1 + \frac{3}{2 k - 1}\,,
\end{equation}
and the metric function $f(r)$ is computed to be \cite{EpM1}
\begin{equation}
f(r) = 1-\frac{\mu}{r^2}+\frac{q}{r^\beta}\,.
\end{equation}
Be aware and notice that the mass $M$ and the electric charge $Q$ of the BH are related to the two parameters $\mu, q$, respectively. 
In particular,  in five dimensions, the parameters $q$ and $C$ are related via \cite{EpM1}
\begin{equation}
q = \alpha  \kappa  \left(-2C^2\right)^k \frac{ (1-2 k)^2 }{3 (k-2)}\,.
\end{equation}
In what follows,  we will consider the case in which the exponent $\beta$ is higher than 2. To guaranties real roots for the metric function $f(r)$, we take the constants $\mu$ and $q$ such that   \cite{EpM1}
\begin{equation}
\mu > 0
\hspace{0.5cm}
\text{and}
\hspace{0.5cm}
0 < q < q_{\text{max}}\,,
\end{equation}
where the upper bound of the charge parameter corresponds to the extremal BHs. Such bound is then given by
\begin{equation}
q_{\text{max}} = \left( \frac{1-2 k}{k-2} \right)
\left(\frac{2-k }{1+k} \ \mu\right)^{\frac{k+1}{2 k-1}}\,.
\end{equation}
Thus, when we take $k=1$, the higher-dimensional version of the Reissner-Nordstr{\"o}m BH \cite{RN} of Maxwell's linear electrodynamics is recovered. Also, the bound for the charge is  $q_{\text{max}} =\mu^2/4$.
In addition, setting $q=0$, the corresponding higher-dimensional version of the Schwarzschild black hole solution  \cite{Tangherlini} is obtained. The thermodynamics of charged black holes with a nonlinear electrodynamics source was studied in Ref. \cite{Gonzalez:2009nn}.

\subsection{Wave equation for scalar perturbations}

In what follows, we will summarize the 
main ingredients to understand the computation of the QN frequencies for scalar perturbations. 
Let us start by considering the propagation of a test scalar field, $\Phi$  in a fixed gravitational background. 
Such field have the following properties:
i) it is assumed to be real, 
ii) it is massless,
iii)  it is electrically neutral, 
and finally
iv) $\Phi$ is non-minimally  coupled to gravity.
Thus, the corresponding action $S[g_{\mu \nu} ,\Phi]$ acquires the simplest form
\begin{align}
S[g_{\mu \nu} ,\Phi] \equiv \frac{1}{2} \int \mathrm{d}^5 x \sqrt{-g}
\Bigl[
\partial^{\mu} \Phi \partial_{\mu} \Phi + \xi R_5 \Phi^2 
\Bigl]\,.
\end{align}
Notice that the parameter $\xi$ control the strength of the non-minimal coupling, and $R_5$ is the Ricci invariant at five dimensions.
Now, we take advantage of the well-known Klein-Gordon equation (see for instance \cite{crispino,Pappas1,Pappas2,Panotopoulos:2019gtn}
and references therein)
\begin{equation}
\frac{1}{\sqrt{-g}} \partial_\mu (\sqrt{-g} g^{\mu \nu} \partial_\nu) \Phi = \xi R_5 \Phi\,.
\end{equation}
In order to decouple and subsequently resolve the Klein-Gordon equation, we take into consideration the symmetries of the metric and propose as ansatz the following separation of variables:
\begin{equation}\label{separable}
\Phi(t,r,\theta,\phi,\psi) = e^{-i \omega t} \: \frac{y(r)}{r^{3/2}} \: \tilde{Y}_l (\Omega)\,,
\end{equation}
where $\omega$ is the unknown frequency (which will be determined), while $\tilde{Y}_l(\Omega)$ is the five-dimensional generalization of the spherical harmonics, and they depend on the angular coordinates only \cite{book}. After the implementation of the above mentioned  ansatz it is easy to obtain, for the radial part, the following equation
\begin{equation}\label{pseudov1}
f(r)^2 y''(r) + f(r) f'(r) y'(r) + \left( \omega^2 - f(r) \left(\xi R_5 + \frac{l (l+2)}{r^2} + \frac{3}{2} \frac{f'(r)}{r} + \frac{3}{4} \frac{f(r)}{r^2} \right) \right) =0 \,,
\end{equation}
where the prime denotes derivative with respect to $r$. Now, changing variable to the the well-known tortoise coordinate $x$, i.e.,
\begin{equation}
x  \equiv  \int \frac{\mathrm{d}r}{f(r)}\,.
\end{equation}
We obtain the Schr{\"o}dinger-like equation, namely
\begin{equation}
\frac{\mathrm{d}^2 y}{\mathrm{d}x^2} + [ \omega^2 - V(x) ] \: y = 0\,,
\end{equation}

\begin{figure}[ht!]
\centering
\includegraphics[scale=0.48]{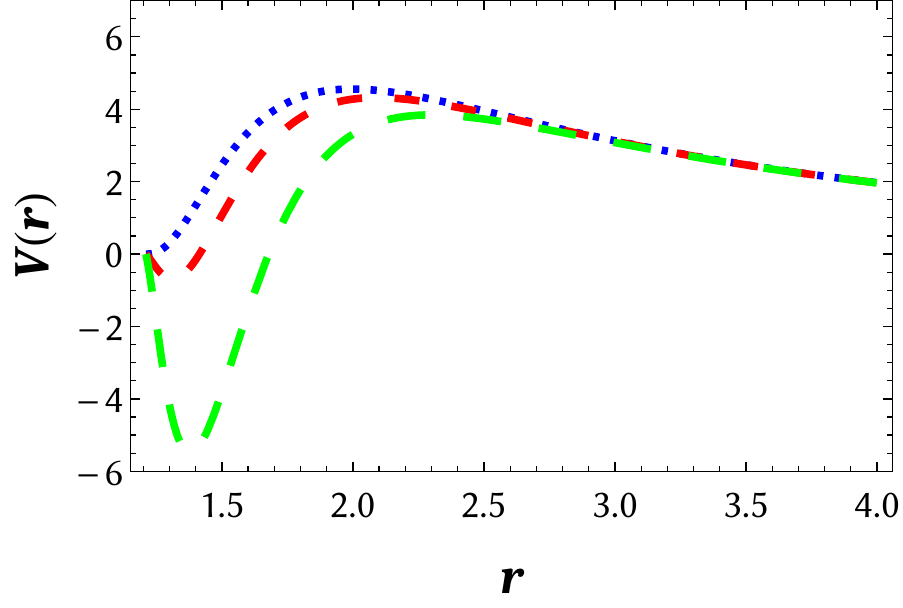} \
\includegraphics[scale=0.48]{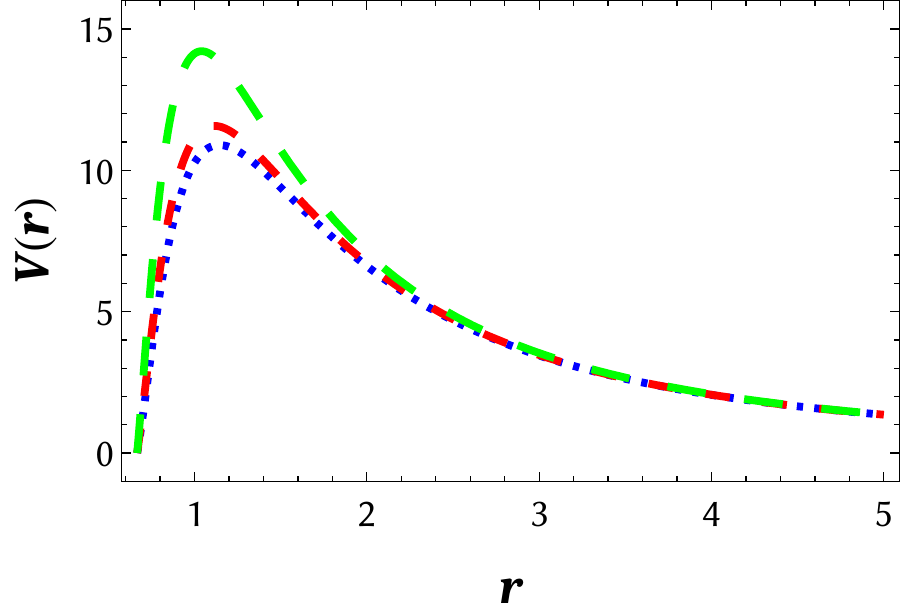} \
\includegraphics[scale=0.48]{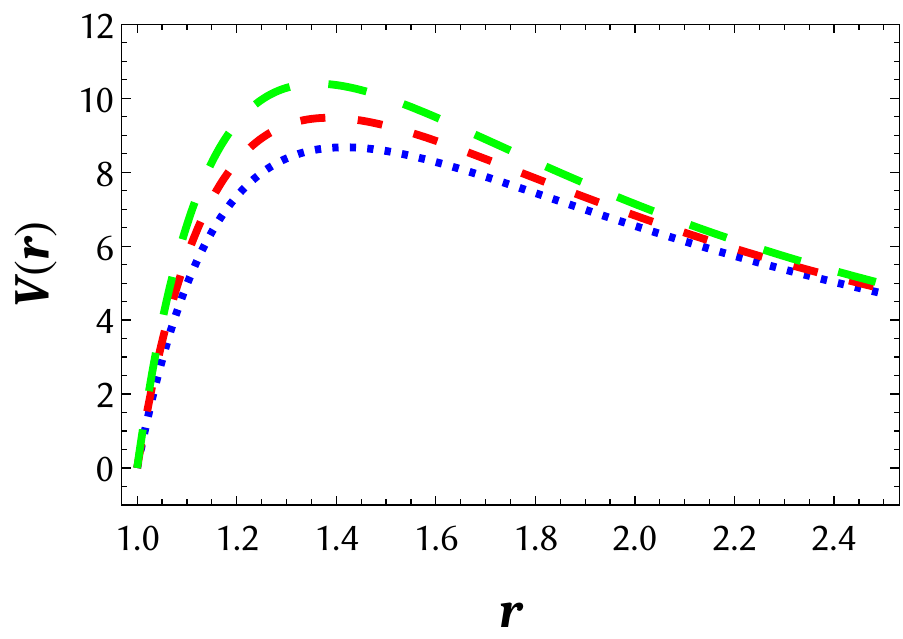}
\caption{
Effective potential for scalar perturbations against the  radial coordinate for $\mu=2$ and $l=5$.
{\bf Left panel:}  Assuming $k = 0.75$ and  $q = 1.39$ we plot three curves:
i) short - dashed blue line for $\xi = 5 $, 
ii) dashed red line for $\xi = 15 $ 
and 
iii) long - dashed green line for $\xi = 50 $.
{\bf Middle panel:} Assuming $k = 1.5$ and $q = 1.27$ we plot three curves:
i) short - dashed blue line for $\xi = 5 $, 
ii) dashed red line for $\xi = 15 $ 
and 
iii) long - dashed green line for $\xi = 50 $.
{\bf Right panel:} Assuming $k = 1.75$ and $q = 1$ we plot three curves:
i) short - dashed blue line for $\xi = 5 $, 
ii) dashed red line for $\xi = 50 $ 
and 
iii) long - dashed green line for $\xi = 100 $.
}
\label{fig:1} 	
\end{figure}

Finally, the effective potential (for scalar perturbations) in five-dimensions is then given by \cite{ref}
\begin{equation}
V(r) = f(r) \: \left(\xi R_5 + \frac{l (l+2)}{r^2} + \frac{3}{2} \: \frac{f'(r)}{r} + \frac{3}{4} \: \frac{f(r)}{r^2} \right)\,.
\end{equation}
As always, the prime denotes differentiation with respect to the radial coordinate, and $l \geq 0$ is the angular degree.
As benchmarks, we will take three concrete values for the exponent $k$  in Fig.~\ref{fig:1}. Note that depending on the value of $k$, the potential can be a potential well, left panel, or a  potential barrier, central and right panel. As we will see, in the next section, the first type of potential lead to unstable modes, while that the second one lead to stable modes.

\clearpage

\section{Quasinormal frequencies}

\subsection{Stable modes}


In what follows, we will compute the corresponding QNFs for $\mu=2$, $l=5$, considering small and large values of the charge for three different values of the exponent $k$, and varying the non-minimal coupling constant $\xi$.
Be aware and notice we have an allowed range for $q$, ranging from zero  to $ q_{\text{max}}$. 
Given that the WKB method is well-known, we take advantage of such a fact to avoid the inclusion of unnecessary details. 
We then show that QN spectra may be computed via the following expression
\begin{equation}
\omega_n^2 = V_0+(-2V_0'')^{1/2} \Lambda(n) - i \nu (-2V_0'')^{1/2} [1+\Omega(n)]\,,
\end{equation}
where 
i) $V_0''$ is the second derivative of the potential evaluated at the maximum, 
ii) $\nu=n+1/2$, $V_0$ is the maximum of the effective potential barrier, 
iii) $n=0,1,2...$ is the overtone number, 
while the functions
$\Lambda(n), \Omega(n)$ are complex expressions of $\nu$ (and derivatives of the potential evaluated at the maximum), and they can be consulted in \cite{paper2,paper7}. 
At this level, should be mentioned that the 3rd order approximation was first constructed by Iyer and Will in Ref. \cite{wkb2} and subsequently generalized. Thus, to perform our computations, we have used
the Wolfram Mathematica \cite{wolfram} code utilizing WKB method at any order from one to six \cite{code}.
In particular, notice that we will use the sixth order approximation. Also, should be mentioned that, for a concrete angular degree $l$, we have considered values $n < l$ only,  see e.g. Tables \eqref{tableset1} to \eqref{tableset6} of the present manuscript (Appendix \ref{Numericalvalues}). For higher order WKB corrections (and recipes for simple, quick, efficient and accurate computations) see \cite{Opala,Konoplya:2019hlu,RefExtra2}. In particular, we should mention that as the WKB series
converges only asymptotically, there is no mathematically strict criterium for evaluation of an error according to \cite{Konoplya:2019hlu}. However, the sixth/seventh order usually produce the best results. In that direction, taking into account the Pad{\'e} approximations we can have a higher accuracy of the WKB
approach, however, this analysis will be performed in a future study.

\begin{figure}[ht!]
\centering
\includegraphics[scale=0.73]{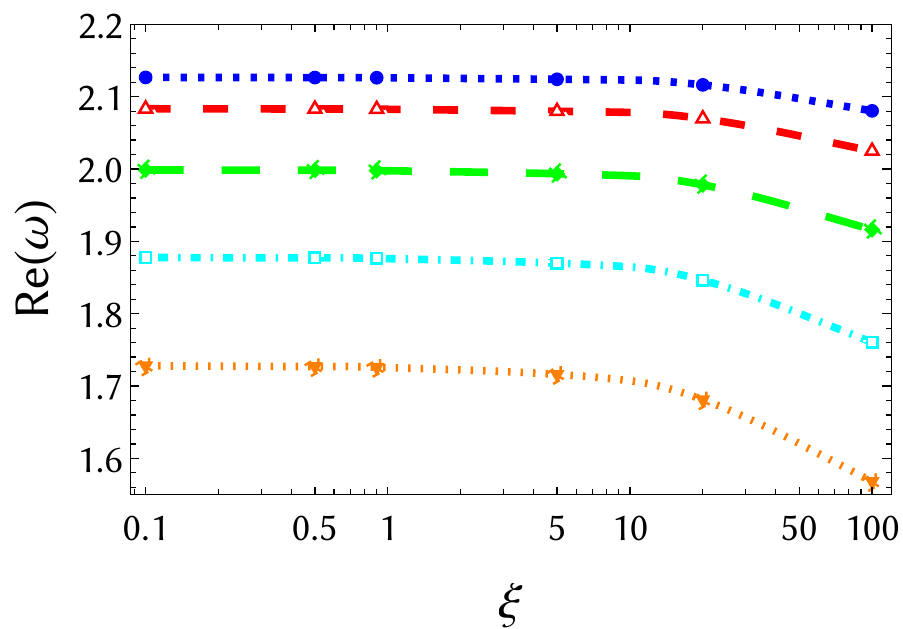} \
\includegraphics[scale=0.73]{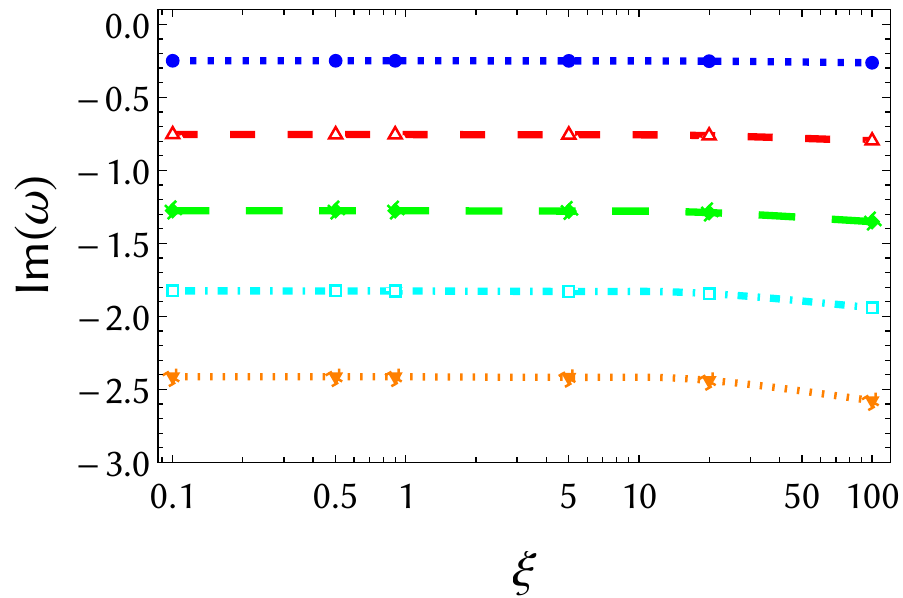}
\\
\includegraphics[scale=0.73]{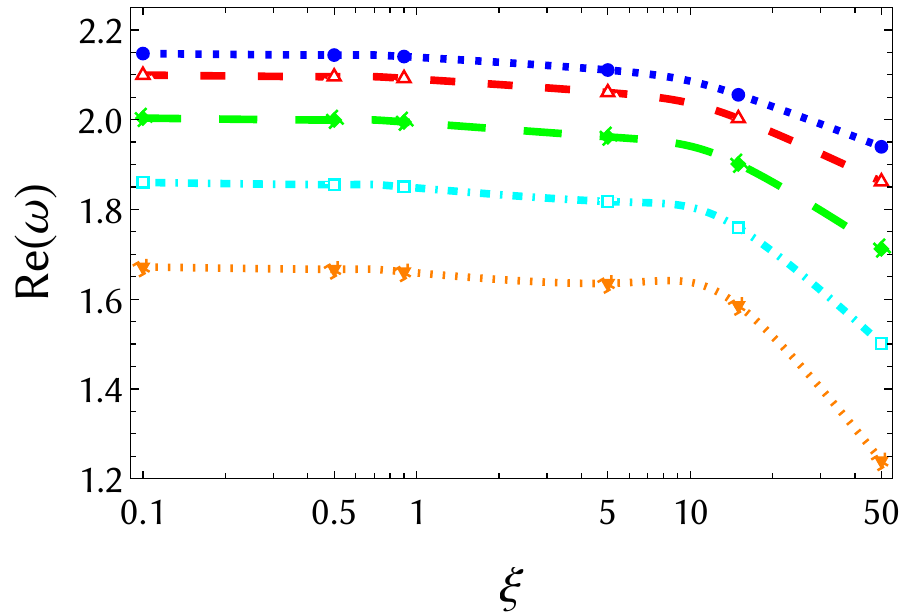} \
\includegraphics[scale=0.73]{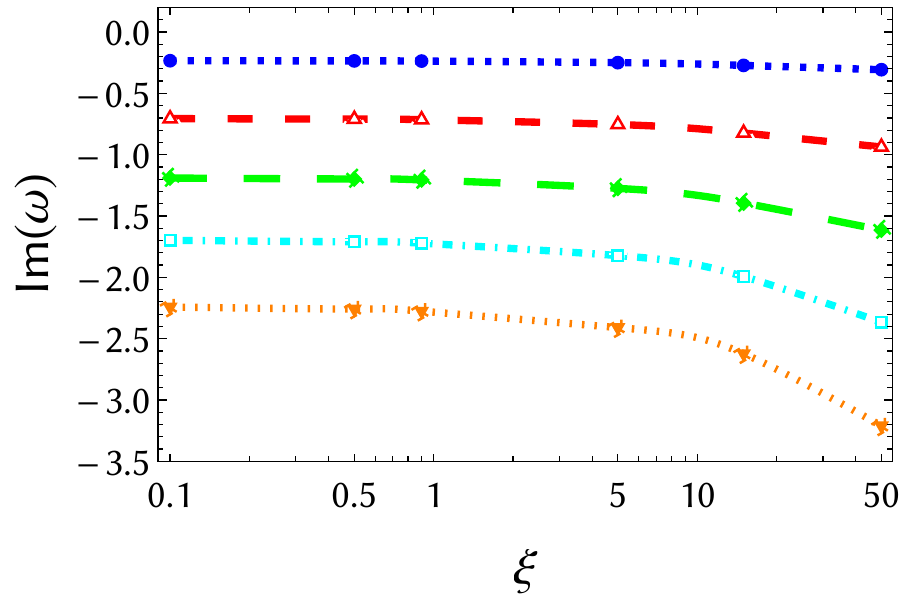}
\caption{
Quasinormal modes (real and imaginary parts) against the non-minimal coupling $\xi$. 
The first row corresponds to the 
sixth table
where the numerical values used are as follows: $q=0.1$, $\mu=2$, $l=5$ and $k=1.75$ for several values of the 
non-minimal coupling parameter $\xi$.
The second row corresponds to 
seventh table
where the numerical values used are as follows: $q \approx 1.39$, $\mu=2$, $l=5$ and $k=0.75$ for several values of 
the non-minimal coupling parameter $\xi$. 
The color codes is as follow: 
i) $n=0$ (short-dashed blue line)
ii) $n=1$ (dashes red line)
iii) $n=2$ (long-dashed green line)
iv) $n=3$ (dot-dashed cyan line)
v) $n=4$ (dotted orange line)
}
\label{fig:2} 	
\end{figure}

\begin{figure}[ht!]
\centering
\includegraphics[scale=0.73]{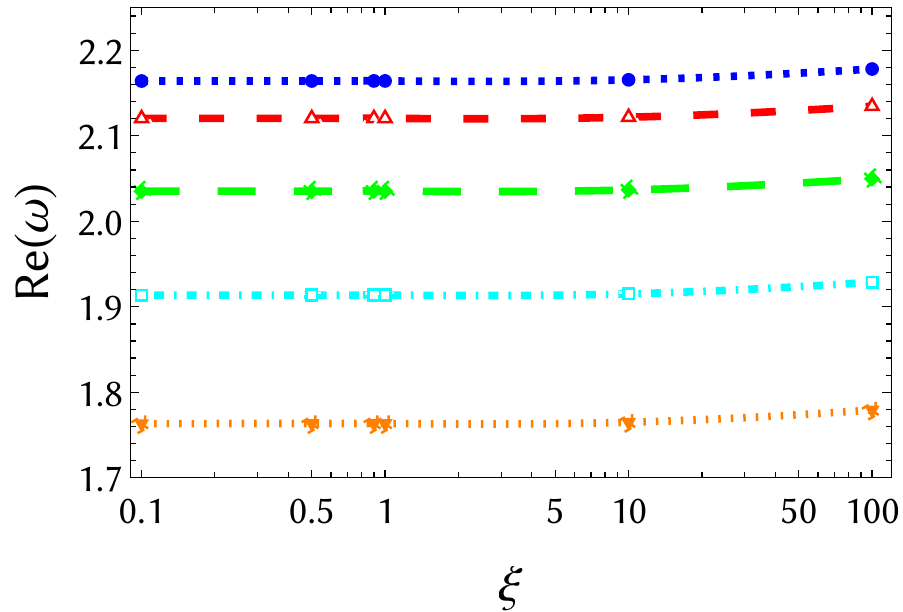} \
\includegraphics[scale=0.73]{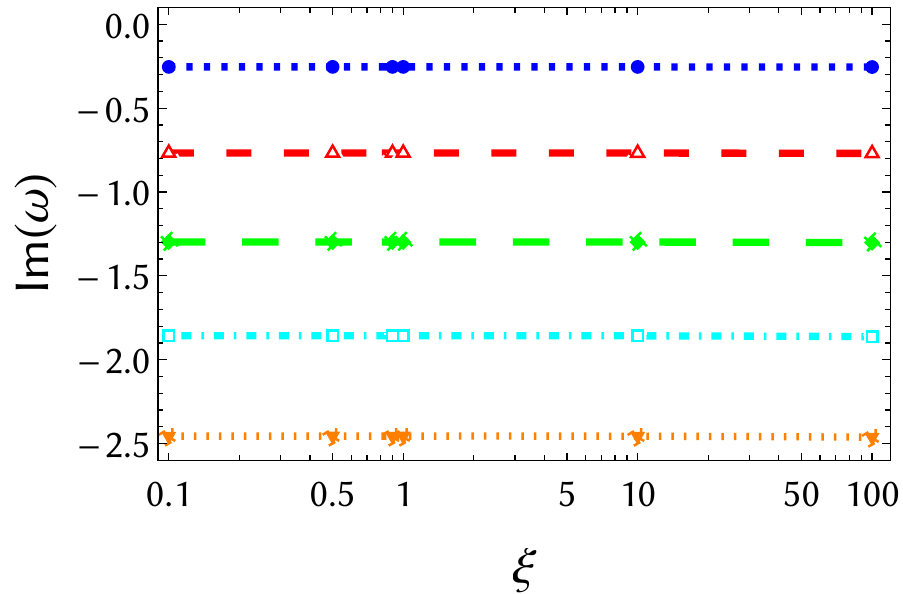}
\\
\includegraphics[scale=0.73]{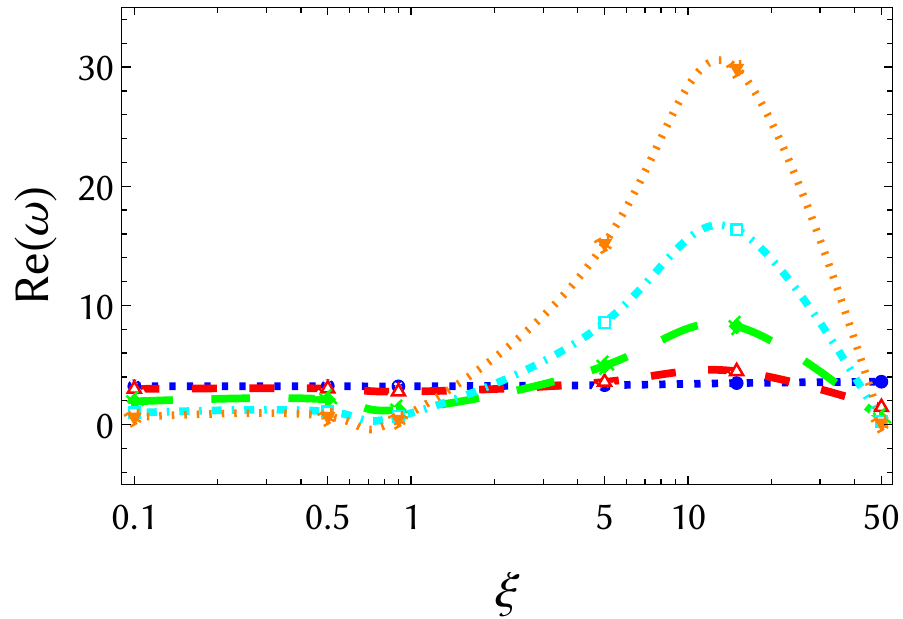} \
\includegraphics[scale=0.73]{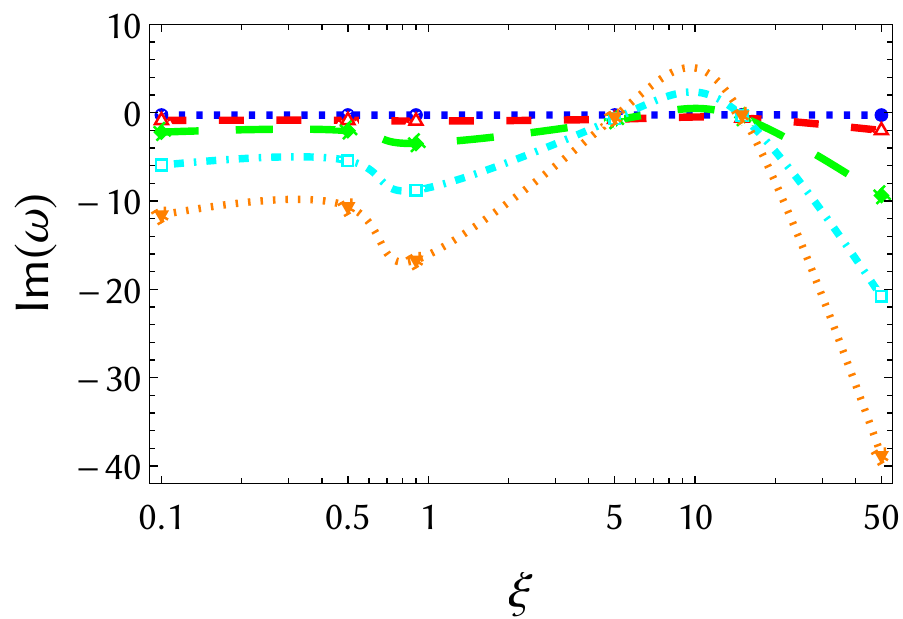}
\caption{
Quasinormal modes (real and imaginary parts) against the non-minimal coupling $\xi$. The first row corresponds to the eighth table
where the numerical values used are as follows: $q=0.1$,  $\mu=2$, $l=5$ and $k=1.5$ for several values of the non-minimal coupling parameter $\xi$.
The second row corresponds to ninth table
where the numerical values used are as follows: $q=1.27$, $\mu=2$, $l=5$ and $k=1.5$ for several values of the non-minimal coupling parameter $\xi$.
The color code is as above.
}
\label{fig:3} 	
\end{figure}

\begin{figure}[ht!]
\centering
\includegraphics[scale=0.73]{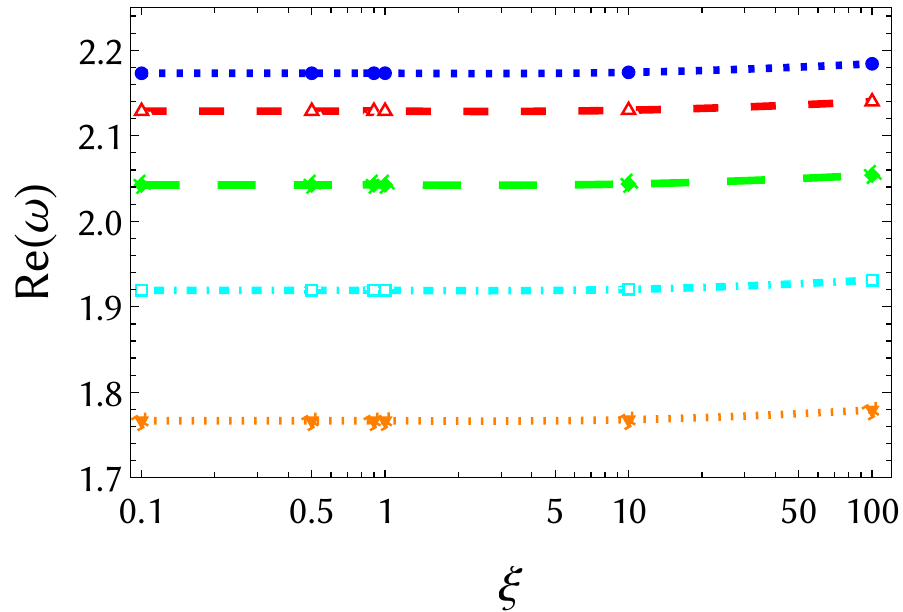} \
\includegraphics[scale=0.73]{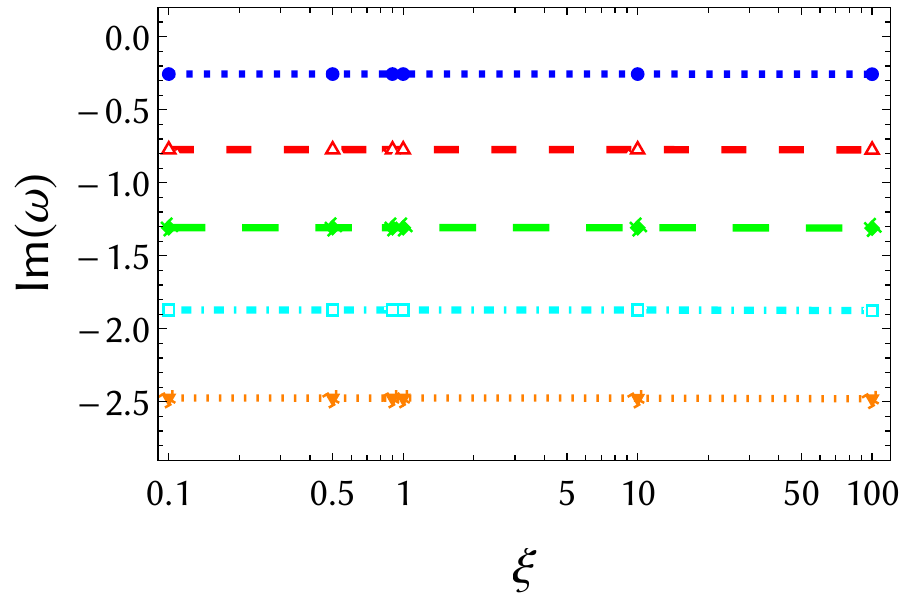}
\\
\includegraphics[scale=0.73]{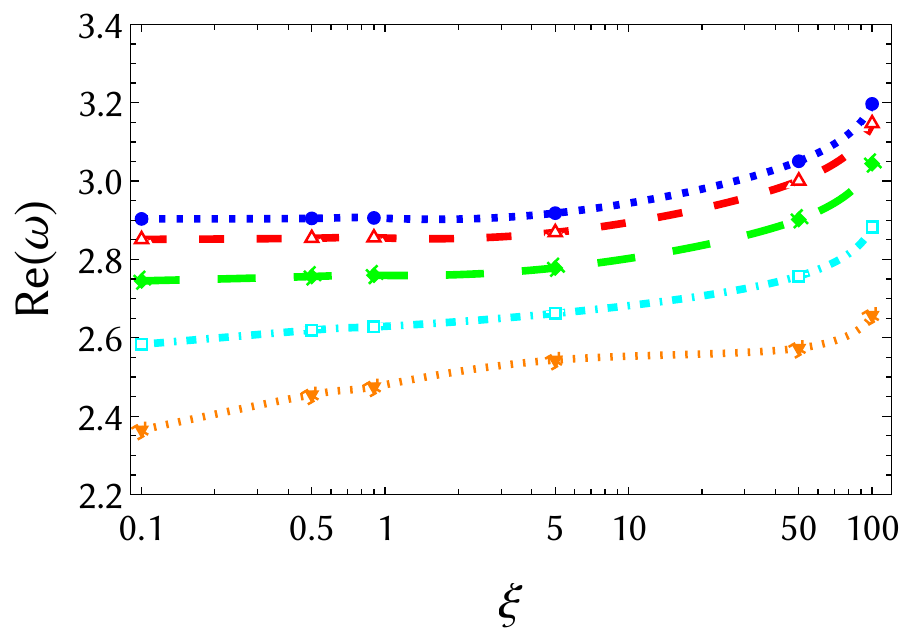} \
\includegraphics[scale=0.73]{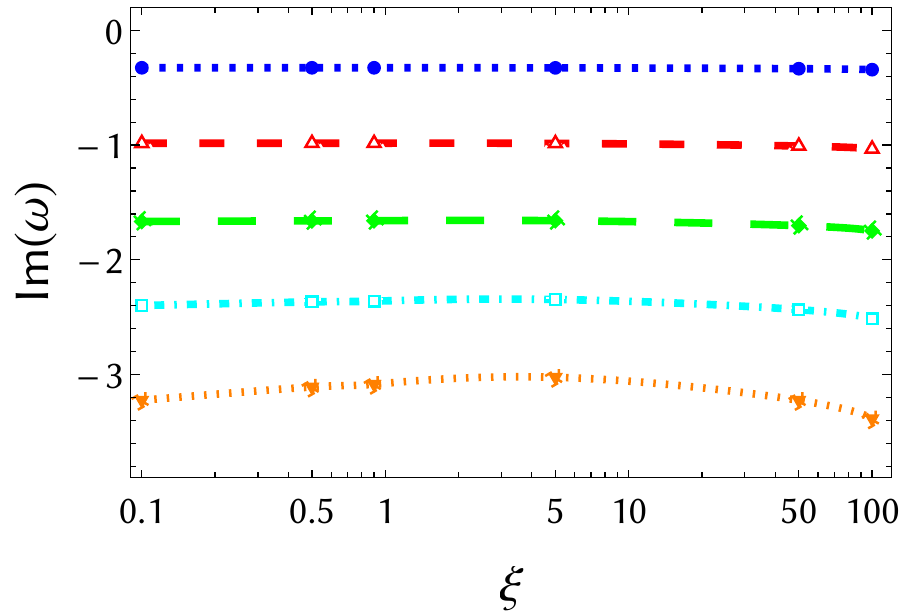}
\caption{
Quasinormal modes (real and imaginary parts) against the non-minimal coupling $\xi$. First row corresponds to the tenth table
where the numerical values used are $q=0.1$, $\mu=2$, $l=5$ and $k=1.75$ for several values of the non-minimal coupling parameter $\xi$.
The second row corresponds to the eleventh  table
where the numerical values used are $q=1$, $\mu=2$, $l=5$ and $k=1.75$ for several values of the non-minimal coupling parameter $\xi$.
The color codes is as above.
}
\label{fig:4} 	
\end{figure}

Our main numerical results are summarized in Fig.~\eqref{fig:2}, \eqref{fig:3} and \eqref{fig:4} as well as in the corresponding tables
~\eqref{tableset1}, \eqref{tableset2}, \eqref{tableset3}, \eqref{tableset4}, \eqref{tableset5}, and \eqref{tableset6}. 
We then show the corresponding QNMs by plotting the real and imaginary parts against the non-minimal coupling $\xi$, for different values of the parameters involved. 
We also show that, for a given set of parameters, when we increase the overtone number the real and imaginary part, in general, decreases.

To summarize, we can say that the spectrum obtained exhibits the following properties:
i) the real part of the frequencies, $\text{Re}(\omega_n)$, is always positive while the imaginary part, $\text{Im}(\omega_n)$, always is negative. In light of the above-mentioned,  all modes are found to be stable, 
ii) the real part decreases with $n$ (except in rare cases), 
iii) the absolute value of the imaginary part increases with $n$ (except in rare cases).




\clearpage

Now, in order to visualize the effect of the angular number $l$ on the QNFs, we show in Tables \ref{table:set_P1} and \ref{table:set_P12} the QNFs for $k=0.75$, and $k=1.50$, respectively, and different values of $\xi$, and $l$. Here, we consider the family of the complex modes. We can observe that the longest-lived modes are the ones with higher angular number for massless scalar field, according with other spacetimes, see \cite{Aragon:2020tvq, Aragon:2020xtm, Aragon:2020teq, Fontana:2020syy}. Also, we show Table \ref{table:set_P1} the QNFs via the WKB and the pseudospectral Chebyshev methods in order to check both methods. 

\smallskip

Let us briefly explain how the pseudospectral Chebyshev method works. First, in Eq. \eqref{pseudov1} we make the change of variable $z = 1-r_H / r$, and the variable $y$ now is a function of $z$. Then we incorporate the boundary conditions, we impose the wave to be purely ingoing at the horizon ($z=0$), and purely outgoing at spatial infinity ($z=1$), therefore we define $y(z) = z^{- i \omega/f'(r_H)} e^{i \omega r_H /(1-z)} \chi (z)$, where $\chi$ satisfies the appropiate boundary conditions to the eigenvalue problem under consideration. Then, the result can be expanded in a complete basis of functions, namely $\{\phi_i(z)\}: \chi(z) = \sum_{i=0}^{\infty}  c_i \phi_i(z)$, where $c_i$ are the coefficients of the expansion, and we choose the Chebyshev polynomials as the complete basis, which are defined by $T_j(x) = \cos(j \cos^{-1}x)$, where $j$ corresponds to the grade of the polynomial. The sum must be truncated until some $N$ value, therefore the function $\chi(z)$ can be approximated by $N$ in such a way that 
$\chi(z) \approx \sum_{i=0}^N c_i T_i(x)$. Given that, by construction $ 0 \leq z \leq 1 $, the connection between the variable $x$ and $z$ is then $x = 2z - 1$. The interval [0,1] is, therefore, discretized at the Chebyshev collocation points $z_j$ by using the so-called Gauss-Lobatto grid, where $z_j = \frac{1}{2}[1 - \cos(\frac{j\pi}{N})]$, with $j =0,1,...,N$. To finish, the differential equation is evaluated at each collocation point. Therefore, a system of $N + 1$ algebraic equations is obtained, and its corresponds to a generalized eigenvalue problem which is solved numerically to obtain the QNMs spectrum, by employing the built-in Eigensystem [ ] procedure in Wolfram’s Mathematica \cite{wolfram}.

\smallskip

For the pseudospectral method, we use a value of $N$ in the interval [95-105] for the majority of the cases
which depends on the convergence of $\omega$ to the desired accuracy.  We will use an accuracy of eight or nine  decimal places for the majority of the cases. In addition, to ensure the accuracy of the results, the code was executed for several increasing values of $N$ stopping when the value of the QNF was unaltered. We can observe a less difference  of the QNFs obtained via both methods when $l$ increases.  As we mentioned, in Table \ref{table:set_P12} we show the QNFs for $k=1.5$, in this case is not possible to find QNFs that converge by using pseudospectral Chebyshev methods, which limit our analysis. So, we can give a deeper analysis, as we will see in the following for $k < 1$, but for $k > 1$, as we have shown, we have obtained only complex frequencies, leaving outside possibles QNFs, as the purely imaginary, that generally appear by using the pseudospectral Chebyshev method, as in other spacetimes.   

\begin{table}[h] 
\centering
\caption{QNFs for scalar perturbations via the WKB and the pseudospectral Chebyshev methods for  $q=0.1$, $\mu=2$, and $k=0.75$ for several values of the non-minimal coupling parameter, $\xi$, and the angular degree $l$. }
\resizebox{\columnwidth}{!}{%
\begin{tabular}{cccc}
\hline
$l$ &  $\xi=0.1$ & $\xi=5$ & $\xi=50$ \\
\hline
  0        &  0.377443 - 0.269536 I  &  0.347908 - 0.266779 I & 0.286247 - 0.459588 I  \\
  0 (WKB)  &  0.384833 - 0.259599 I        &  0.345895 - 0.217895 I       & 0.284643 - 0.356436 I        \\
  1        &  0.718764 - 0.255032 I  & 0.706724 - 0.257024 I  & 0.632025 - 0.313644 I  \\
  1 (WKB)  &  0.717699 - 0.257415 I        & 0.704157 - 0.25834 I         & 0.644719 - 0.3147 I       \\
  5        & 2.126587 - 0.249752 I     & 2.123992 - 0.250408 I    & 2.101789 - 0.256645 I \\
  5  (WKB) & 2.12659 - 0.249749 I          & 2.12399 - 0.250399 I         & 2.10179 - 0.256661 I      \\
  20       & 7.431488 - 0.249136 I     & 7.430828 - 0.249194 I    & 7.424795 - 0.249734 I \\
  20 (WKB) & 7.43149 - 0.249136 I          & 7.43083 - 0.249194 I         & 7.4248 - 0.249734 I       \\
\hline   

\end{tabular}
}
\label{table:set_P1}
\end{table}




\begin{table}[h] 
\centering
\caption{QNFs for scalar perturbations for  $q=0.1$, $\mu=2$, and $k=1.5$ for several values of the non-minimal coupling parameter, $\xi$, and the angular degree, $l$, using the WKB method. }
\resizebox{\columnwidth}{!}{%
\begin{tabular}{cccc}
\hline
$l$ &  $\xi=0.1$ & $\xi=5$ & $\xi=50$ \\
\hline
  0  & 0.392846\, - 0.259043 I        &  0.398241\, - 0.259198 I      & 0.432526\, - 0.270273 I       \\
  1  &  0.730638\, - 0.261403 I        & 0.732645\, - 0.261827 I        & 0.751854\, - 0.265274 I       \\
  5  & 2.16401\, - 0.253957 I          & 2.1647\, - 0.253996 I        & 2.17103\, - 0.254357 I    \\
  20 & 7.56188\, - 0.253307 I          & 7.56207\, - 0.25331 I       & 7.56387\, - 0.253338 I      \\
\hline   
\end{tabular}
}
\label{table:set_P12}
\end{table}


\subsection{Unstable modes}

As was already mentioned, the existence of a potential well is possible, see left panel Fig. \ref{fig:1}, which depends on the value of the parameter $k$, and there are bound states for massless scalar fields which allows to accumulate the energy to trigger the instability. However, the potential well has to be deep enough in order to guarantee the stability of those bound states, due to the zero point energy of quantum systems \cite{Grain:2007gn}. Thus, as in Refs. \cite{Bronnikov:2012ch, Liu:2020evp} we can view the existence of a negative potential well as a necessary but not sufficient condition for the instability. 
A potential well exists in the vicinity of the event horizon in our system when $V_{eff}'(r_{H}) < 0$  as in Ref. \cite{Gonzalez:2022upu}, which yields
\begin{equation}\label{condt}
-2 \xi  r_H^2 f''(r_H)+3 (1-4 \xi ) r_H f'(r_H)+2 l (l+2)+12 \xi 
< 0\,,
\end{equation}

and unstable modes would appear for a deep enough potential well. When the above condition is not satisfied for some set of parameters, the potential well disappears and the background becomes stable under scalar perturbations because the perturbation can be easily absorbed by the black hole.

In order to consider more complicated potentials. In Tables \ref{Table1}, \ref{Table2} and \ref{Table3} we consider a  potential
where the condition (\ref{condt}) is not violated, and we can observe that a transition from stable and oscillatory QNFs to purely imaginary and unstable QNFs, when $r_H$ decreases. However, in Table \ref{Table4} the condition (\ref{condt}) is  violated, for small values of $\xi<10$, and $r_H$ is constant. The QNFs are oscillatory and stable. While that for $\xi>10$, the condition (\ref{condt}) is not violated, and there is a transition from oscillatory and stable QNFs to purely imaginary and unstable QNFs, when the value of the parameter $\xi$ increases. With the data summarized in Tables \ref{Table1}, \ref{Table2}, \ref{Table3} and \ref{Table4} we present the parametric region of instability from the numeric point of view and with the Eq. (\ref{condt}) we characterized the necessary condition for the existence of a potential well.

\begin{table}[H] 
\centering
\caption{Fundamental QNFs for massless scalar perturbations for $\mu=2$, $\xi=50$, $k=0.75$ and $l=5$, for several values of $q$. Here, we have used the pseudospectral Chebyshev method.}
\resizebox{1\columnwidth}{!}{%
\begin{tabular}{ccccccc}
\hline
$q$ &   $0$ &  $0.1$ & $0.2$ &   $0.3$ &   $0.4$  &  $0.5$ \\
\hline
 $\omega$   &  2.12502086-0.25070686 I  & 2.10178921-0.25664532 I   & 2.08151163-0.26301349 I & 2.06374377-0.26907083 I  & 2.04784247-0.27451745 I & -0.171610883 I   \\
 \hline
 $q$ & $0.55$  &   $0.6$ &  $0.7$ & $0.8$ &   $0.9$ &   $1$  \\
\hline
  $\omega$ &  -0.01043062933 I &  0.13903435 I  &  0.41237443 I   &  0.66263041 I   &  0.89911396 I   &  1.12875673 I    \\   
\hline
\end{tabular}
}
\label{Table1}
\end{table}

\begin{table}[H] 
\centering
\caption{Fundamental QNFs for massless scalar perturbations for $\mu=2$, $\xi=50$, $q=1$ and $l=5$, for several values of $k$. Here, we have used the pseudospectral Chebyshev method.}
\resizebox{1\columnwidth}{!}{%
\begin{tabular}{ccccccc}
\hline
$k$ &   $0.55$ &  $0.6$ & $0.62$ &   $0.64$ &   $0.65$  &  $0.67$ \\
\hline
  $\omega$  & 2.12517781-0.25067652 I  & 2.12438790-0.26526377 I  & -0.20138421 I  &  0.18315691 I   &  0.33864551 I   & 0.59293385 I  \\
 \hline
 $k$ &   $0.7$ &  $0.72$ & $0.75$ &   $0.77$ &   $0.8$  &  $0.82$ \\
\hline
  $\omega$  &  0.86635211 I &   0.99504189 I  &  1.12875673 I   &   1.18669872 I   &  1.23665588 I  & 1.24921455 I   \\   
  \hline
\end{tabular}
}
\label{Table2}
\end{table}

\begin{table}[H] 
\centering
\caption{Fundamental QNFs for massless scalar perturbations for $q=1$, $k=0.75$, $\xi=50$ and $l=5$, for several values of $\mu$. Here, we have used the pseudospectral Chebyshev method.}
\resizebox{1\columnwidth}{!}{%
\begin{tabular}{ccccccc}
\hline
$\mu$ & $2.2$ &  $2.4$ & $2.6$ & $2.8$ &   $3$ & $3.2$  \\
\hline
 $\omega$   &  0.43342701 I  & -0.07195799 I &  1.79618206-0.24073196 I &    1.74305536-0.22779439 I &  1.69324593-0.21687689 I  &  1.64662973-0.20758148 I    \\
 \hline
\end{tabular}
}
\label{Table3}
\end{table}

\begin{table}[H] 
\centering
\caption{Fundamental QNFs for massless scalar perturbations for $\mu=2$, $q=1$, $k=0.75$ and $l=5$, for several values of $\xi$. Here, we have used the pseudospectral Chebyshev method.}
\resizebox{1\columnwidth}{!}{%
\begin{tabular}{ccccccc}
\hline
$\xi$ &   $10$ & $15$ &  $20$ & $25$ & $30$ &   $35$   \\
\hline
 $\omega$   &  2.09163108-0.25829400 I  & 2.07174879-0.26594160 I &  -0.02654658 I  & 0.23692838 I  &  0.45959635 I  &  0.65381000 I    \\
 \hline
\end{tabular}
}
\label{Table4}
\end{table}

\begin{figure}[h]
\begin{center}
\includegraphics[width=0.7\textwidth]{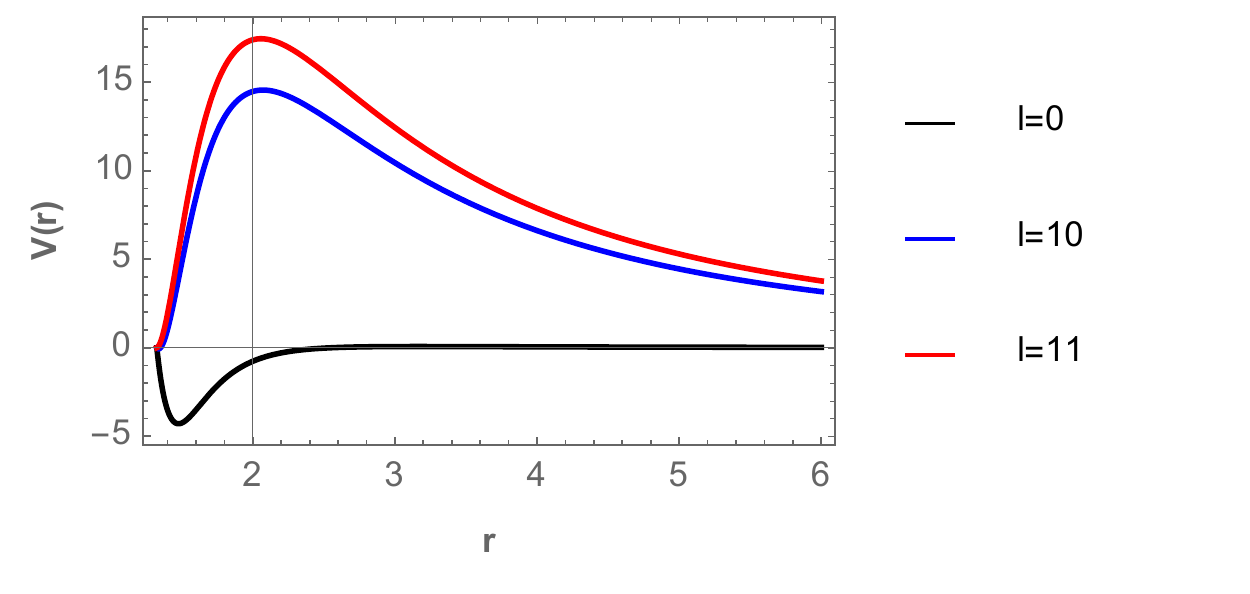}
\includegraphics[width=0.7\textwidth]{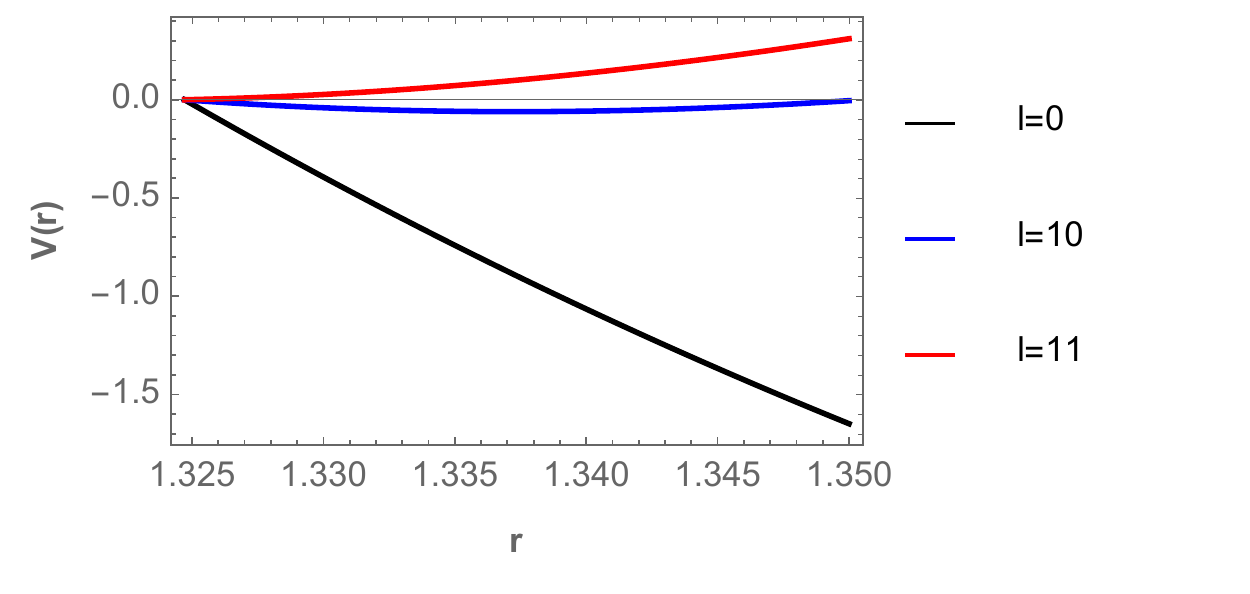}\\
\end{center}
\caption{Effective potential $V(r)$ as a function of $r$, with $\mu=2$, $q=1$, $l=0, 10, 11$, $\xi=50$, and $\kappa=0.75$. Top panel for $r_H<r<6$, bottom panel for $r_H<r<1.35$.}
\label{unstableplot}
\end{figure}

Furthermore, if we consider $l=0, 10, 11$, and fixed values of $\mu, q, l, \xi, \kappa$ in order to observe the behaviour of the modes near the threshold of instability, in Fig. \ref{unstableplot} we show the associated potential. For these values, Eq. (\ref{condt}) yields that the slope of the potential vanishes for $ l \approx 10.78$, but $l$ must be an integer number. So, for $l \leq 10$ there is a potential well  outside the event horizon and for $l>10$ there is a  potential barrier. In Table \ref{Answertable} we show the QNFs. Here, it is possible to observe the appearing of unstable modes for small values of $l$ ($l<10$), where there  is a potential well. Then, a stable and purely imaginary mode  near the threshold of instability for $l=10$, where 
there is a potential well, but one could to say that the potential does not allow bound states, then the scalar field can not to accumulate the energy to trigger the instability. Finally,
for $l>10$ the modes are stable and complex where there is a potential barrier. Therefore, for the case showed in Table \ref{Answertable}, as for some values of multipoles $l$ the QNFs are unstable then the propagation of the scalar field is unstable.

\begin{table}[H] 
\centering
\caption{Fundamental QNFs for massless scalar perturbations using the pseudospectral Chebyshev method for  $q=1$, $\mu=2$, $\xi=50$, and $k=0.75$, for several values of $l$. Here, $r_H\approx 1.325$, and $r_-= 1$.}
\resizebox{1\columnwidth}{!}{%
\begin{tabular}{cccccc}
\hline
$l$ &   $0$ &  $9$ & $10$ &   $11$ &   $20$ \\
\hline
  $\omega$   &   1.75679814 I  & 0.0852972116 I & -0.231851685 I    & 4.16523558 - 0.26181366 I  & 7.41598416 - 0.24738691 I       \\
 \hline
\end{tabular}
}
\label{Answertable}
\end{table}

\newpage

\begin{figure}[h]
\begin{center}
\includegraphics[width=0.42\textwidth]{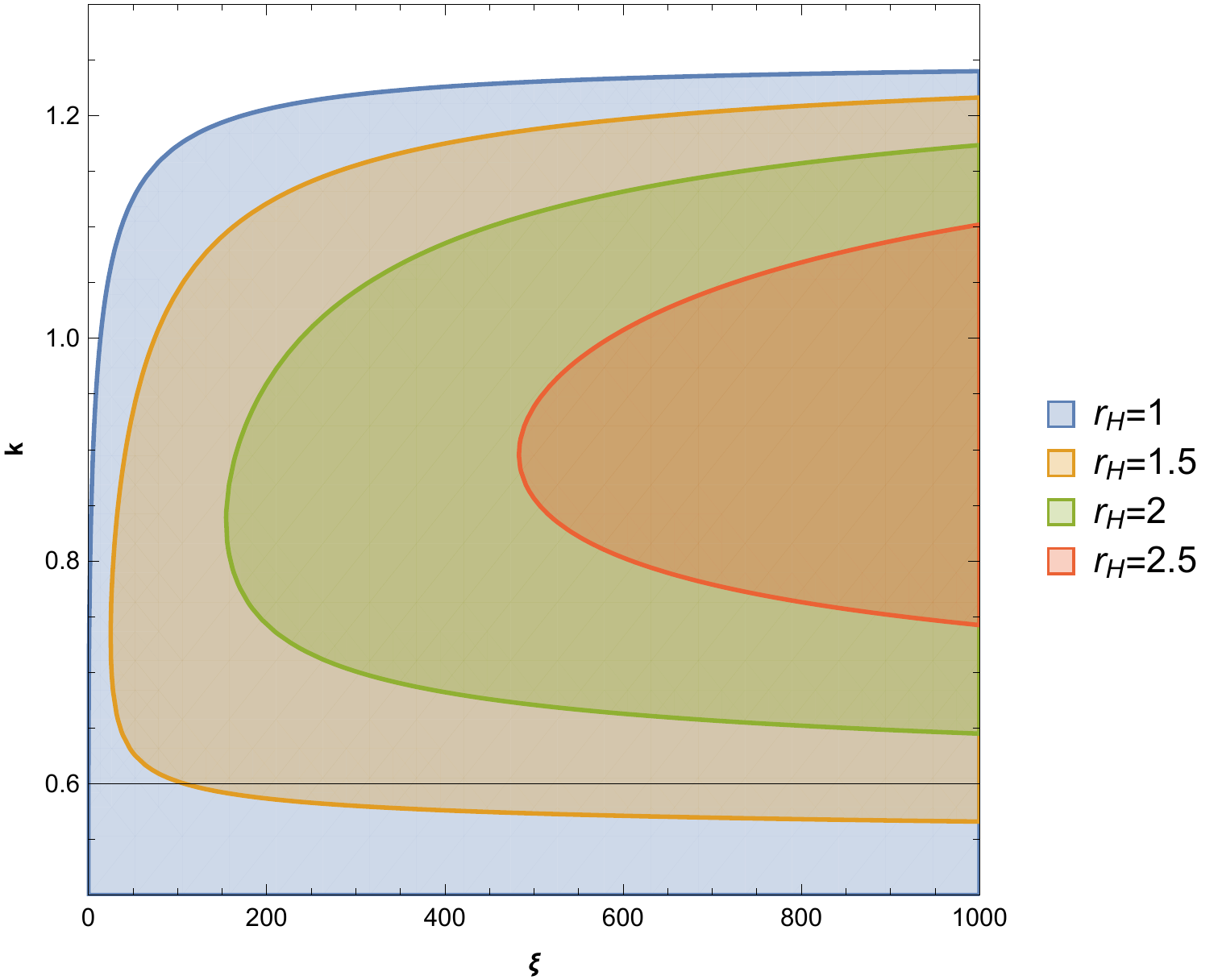}
\includegraphics[width=0.4\textwidth]{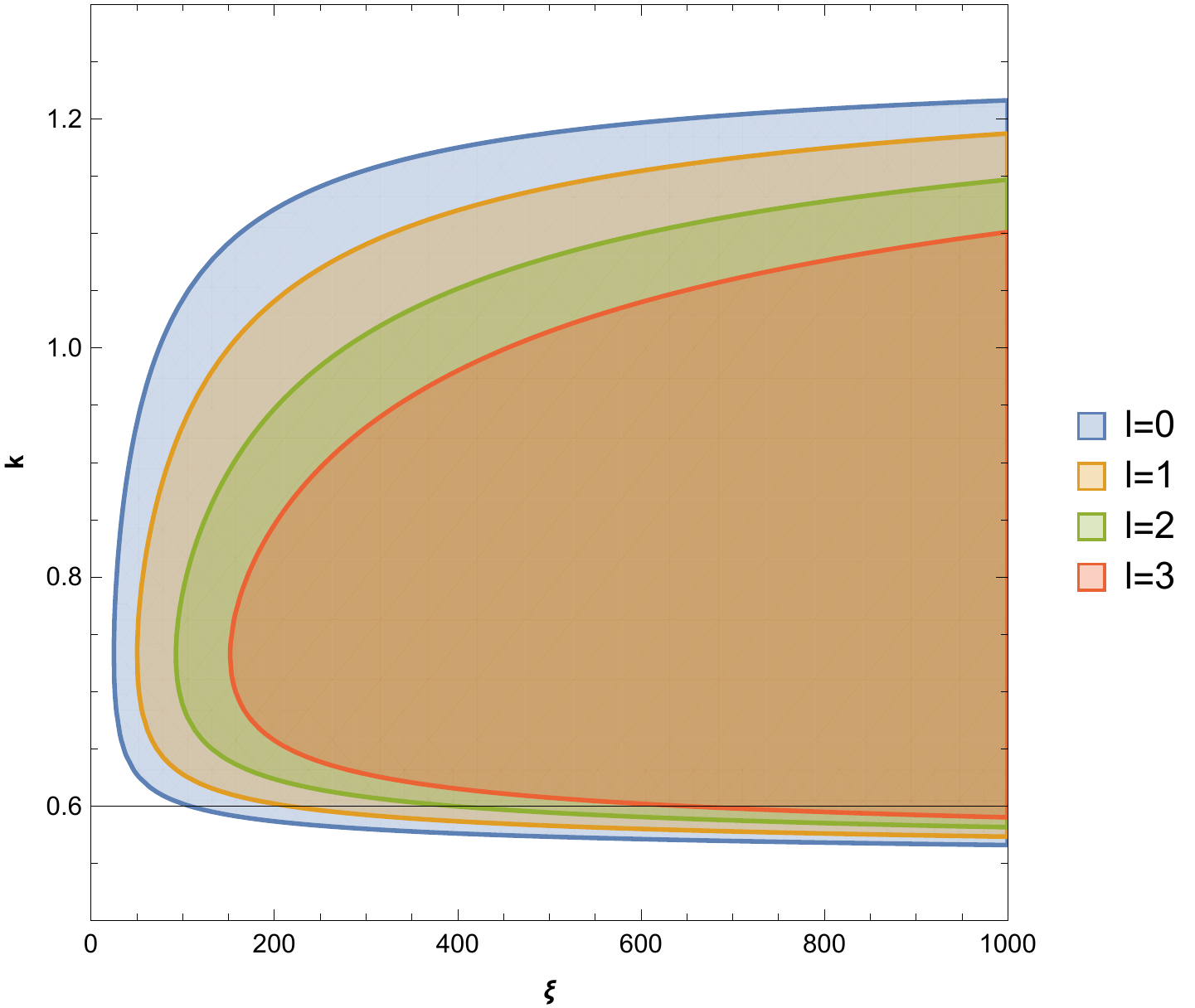}
\end{center}
\caption{We show the parameters space for condition (\ref{condt}). Left panel for different values of the event horizon $r_H$, with $q=0.1$, and $l=0$. Right panel for different values of $l$, with $q=0.1$, and $r_H=1.5$.}
\label{Answerunstableplot2}
\end{figure}

On the other hand, we can observe that the  potential well
depends on the size of the black hole, and the parameters $\xi$ and $k$ as well. We summarize our results in Fig. \ref{Answerunstableplot2}, where we show colored regions where the condition (\ref{condt}) over the plane $(k,\xi)$ is satisfied. In this figure we can see that the area of the region 
decreases when the event horizon increases (left panel), as well as, when $l$ increases (right panel).

It is worth mentioning that for charged spacetimes, the propagation of uncharged scalar fields is described by the existence of two distinct families of modes, one of them are the complex ones associated with the photon sphere, and the other family consist of purely imaginary frequencies which appear when the inner and outer horizon get close, near extremal modes, such transition from complex to purely imaginary dominant modes, as the charge increases, has been found in various charged spacetimes. Also, when the scalar field is charged, the QNMs  become complex \cite{Richartz:2014jla, Richartz:2015saa, Panotopoulos:2019tyg, Cardoso:2017soq, Berti:2003ud, Cardoso:2018nvb, Destounis:2018qnb, Liu:2019lon, Destounis:2019hca, Destounis:2019omd, Destounis:2020pjk, Destounis:2020yav, Aragon:2021ogo, Gonzalez:2021vwp}. Here, we can observe a transition from a dominance of stable modes, see Table \ref{table:set_P1}, to unstable modes at the near 
extremal limit that depend on  $l$, see Table \ref{unstable}, 
for low values of $l$. For higher values of $l$ the stables modes are dominant. \\

\begin{table}[h] 
\centering
\caption{Fundamental QNFs for massless scalar perturbations using the pseudospectral Chebyshev method for  $q=1.39$, $\mu=2$, $\xi=50$, and $k=0.75$, for several values of $l$. Here, $r_H\approx 1.210$, and $r_-\approx 1.180$.}
\resizebox{1\columnwidth}{!}{%
\begin{tabular}{ccccc}
\hline
$l$ &   $0$ &   $1$ &   $5$ &   $20$  \\
\hline
  $\omega$   &   0.190553109 I  & 0.156785407 I    & 0.1278929823 I & 1.93954264 - 0.30711773 I    \\
\hline   
\end{tabular}
}
\label{unstable}
\end{table}

\newpage


\section{Conclusions}

In summary, in this work we computed the quasinormal spectrum for scalar  perturbations of five-dimensional charged black holes in the Einstein-power-Maxwell non-linear electrodynamics. The test field that perturbs the gravitational background is taken to be a real, massless,
electrically neutral canonical scalar field, and we have adopted the popular and widely used WKB semi-analytical method, which has been of utility for determine the stable QNFs and the pseudospectral Chebyshev method, when the WKB method does not allow to obtain the QNFs, and which has been of utility for obtain unstable QNFs. \\


Also, two different types of potentials, were described, one of them is a potential well, where we showed that it is a necessary condition, but it is not enough to have unstable modes. Also, we showed the regions in the parameters space $(\xi, k)$, where there is a potential well, and the area of the regions on the $(\xi, k)$ plane decreases when the event horizon increases and when $l$ increases.
The other potential corresponds to a potential barrier, where the QNFs are complex and the modes are stable. Also, for the complex modes we showed that the longest-lived modes are the ones with higher angular number.\\

It could be interesting to consider other method that allow to obtain the QNFs for $k>1$ in order to perform a deeper analysis of the stable modes, by distinguishing the family of modes, and the propagation of massive scalar field, we hope to address it, in a forthcoming work. Probably, in this case, one could distinguish the family of modes, their dominance, and also to find a critical mass where beyond this values the longest-lived modes are the ones with smaller angular number. Also, it would be interesting to study the quasinormal  spectra for other types of fields such as Dirac or electromagnetic fields and gravitational perturbations.


\section*{Acknowlegements}

We thank the referee for his/her valuable comments and suggestions. This work is partially supported by ANID Chile through FONDECYT Grant Nº 1220871  (P.A.G., and Y. V.), and Grant Nº 1210635 (J. S.).
The author A.~R. acknowledges Universidad de Tarapac\'a for financial support. 


\clearpage

\appendix{}


\section{Numerical values}
\label{Numericalvalues}

\begin{table*}[h] 
\centering
\caption{QNFs for scalar perturbations via the WKB method
for  $q=0.1$, $\mu=2$, $l=5$ and $k=0.75$ for several values of the non-minimal coupling parameter $\xi$. }
\resizebox{0.8\columnwidth}{!}{%
\begin{tabular}{cccc}
\hline
$n$ &  $\xi=0.1$ & $\xi=0.5$ & $\xi=0.9$ \\
\hline
  0   &  2.12659 - 0.249749 I  &  2.12638 - 0.249801 I & 2.12616 - 0.249854 I  \\
    1  &  2.08323 - 0.754668 I & 2.08295 - 0.754819 I & 2.08267 - 0.754971 I  \\
   2  & 1.99865 - 1.27612 I & 1.99823 - 1.27635 I & 1.9978 - 1.27658 I \\
   3  & 1.87772 - 1.82536 I & 1.87708 - 1.82563 I & 1.87643 - 1.82593 I  \\
   4  & 1.72787 - 2.4128 I & 1.72693 - 2.41309 I & 1.72596 - 2.41342 I  \\
\hline   
\hline
$n$ &  $\xi=5$ & $\xi=20$ & $\xi=100$ \\
\hline
   0   &  2.12399 - 0.250399 I  &  2.11627 - 0.252445 I & 2.08033 - 0.263545 I  \\
    1  & 2.0798 - 0.756553 I & 2.06959 - 0.762663 I & 2.02484 - 0.797258 I  \\
   2  & 1.99348 - 1.27906 I & 1.97808 - 1.28931 I & 1.9161 - 1.35156 I \\
   3  & 1.86981 - 1.82914 I & 1.8462 - 1.84408 I & 1.76043 - 1.94076 I  \\
   4  & 1.71609 - 2.41729 I & 1.68111 - 2.43839 I & 1.56805 - 2.57681 I  \\
\hline   
\end{tabular}
}
\label{tableset1}
\end{table*}

\begin{table*} 
\centering
\caption{QNFs for scalar perturbations via the WKB method
for  $q \approx 1.39$, $\mu=2$, $l=5$ and $k=0.75$ for several values of the non-minimal coupling parameter $\xi$. }
\resizebox{0.8\columnwidth}{!}{
\begin{tabular}{cccc}
\hline
$n$ &  $\xi=0.1$ & $\xi=0.5$ & $\xi=0.9$ \\
\hline
   0  &  2.14733 - 0.2339 I  &  2.14397 - 0.23532 I & 2.14067 - 0.236719 I  \\
   1  &  2.09943 - 0.70572 I & 2.09571 - 0.71009 I & 2.09209 - 0.714405 I  \\
   2  &  2.0035 - 1.19032 I & 1.99916 - 1.19799 I & 1.99495 - 1.20563 I \\
   3  & 1.86014 - 1.69827 I & 1.85516 - 1.70986 I & 1.85022 - 1.72163 I  \\
   4  & 1.67177 - 2.24214 I & 1.66652 - 2.25841 I & 1.66105 - 2.27566 I  \\
\hline   
\hline
$n$ &  $\xi=5$ & $\xi=15$ & $\xi=50$ \\
\hline
   0  &  2.11065 - 0.249681 I  &  2.05538 - 0.272132 I & 1.93953 - 0.307151 I  \\
   1  & 2.06059 - 0.75416 I & 2.00273 - 0.8233 I & 1.86192 - 0.937341 I  \\
   2  & 1.96171 - 1.27461 I & 1.90129 - 1.39425 I & 1.71218 - 1.61458 I \\
   3  & 1.8177 - 1.82274 I & 1.75914 - 1.99365 I & 1.50113 - 2.36576 I  \\
   4  & 1.6352 - 2.40979 I & 1.58606 - 2.62226 I & 1.23978 - 3.21111 I  \\
\hline   
\end{tabular}
}
\label{tableset2}
\end{table*}

\begin{table*} 
\centering
\caption{QNFs for scalar perturbations via the WKB method
for $q=0.1$,  $\mu=2$, $l=5$ and $k=1.5$ for several values of the non-minimal coupling parameter $\xi$. }
\resizebox{0.8\columnwidth}{!}{%
\begin{tabular}{cccc}
\hline
$n$ &  $\xi=0.1$ & $\xi=0.5$ & $\xi=0.9$ \\
\hline
   0  &  2.16401 - 0.253957 I  &  2.16407 - 0.25396 I & 2.16412 - 0.253964 I  \\
   1  &  2.12032 - 0.76746 I & 2.12038 - 0.767469 I & 2.12043 - 0.767479 I  \\
   2  & 2.03517 - 1.29800 I & 2.03524 - 1.29801 I & 2.03528 - 1.29803 I \\
   3  & 1.91361 - 1.85716 I & 1.91372 - 1.85714 I & 1.91374 - 1.85720 I  \\
   4  & 1.76335 - 2.45565 I & 1.76352 - 2.45553 I & 1.76348 - 2.45569 I  \\
\hline   
\hline
$n$ &  $\xi=1$ & $\xi=10$ & $\xi=100$ \\
\hline
   0   &  2.16414 - 0.253964 I  &  2.1654 - 0.254036 I & 2.17806 - 0.254764 I  \\
    1  & 2.12045 - 0.767481 I & 2.12173 - 0.767695 I & 2.13454 - 0.76986 I  \\
   2  & 2.03531 - 1.29803 I & 2.03661 - 1.29839 I & 2.04974 - 1.30192 I \\
   3  & 1.91377 - 1.85719 I & 1.91512 - 1.85767 I & 1.92874 - 1.86242 I  \\
   4  & 1.76355 - 2.45562 I & 1.76494 - 2.45625 I & 1.77924 - 2.46192 I  \\
\hline   
\end{tabular}
}
\label{tableset3}
\end{table*}

\begin{table*} 
\centering
\caption{QNFs for scalar perturbations via the WKB method
for $q=1.27$, $\mu=2$, $l=5$ and $k=1.5$ for several values of the non-minimal coupling parameter $\xi$. }
\resizebox{0.8\columnwidth}{!}{%

\begin{tabular}{ccccc}
\hline
$n$ &  $\xi=0.1$ & $\xi=0.5$ & $\xi=0.9$ \\
\hline
0  &  3.21532 - 0.27333 I  &  3.22127 - 0.273044 I & 3.20562 - 0.274672 I  \\
1  &  2.98958 - 0.87490 I & 3.02224 - 0.866361 I & 2.76589 - 0.948612 I  \\
2  &  1.94728 - 2.20243 I & 2.1039 - 2.04232 I & 1.24019 - 3.48308 I \\
3  & 0.986473 - 5.93414 I & 1.08297 - 5.42664 I & 0.678619 - 8.77654 I \\
 4  & 0.627977 - 11.5661 I & 0.688916 - 10.6304 I & 0.452687 - 16.6763 I  \\
\hline   
\hline
$n$ &  $\xi=5$ & $\xi=15$ & $\xi=50$ \\
\hline
0  &  3.30853 - 0.267911 I  &  3.48324 - 0.26023 I & 3.59523 - 0.277533 I  \\
1  & 3.57635 - 0.736656 I & 4.46767 - 0.604084 I & 1.45626 - 2.0183 I  \\
2  & 4.92945 - 0.87228 I & 8.25419 - 0.536388 I & 0.49512 - 9.38322 I \\
3  & 8.56712 - 0.676706 I & 16.3517 - 0.369498 I & 0.27506 - 20.7987 I  \\
4  & 15.1886 - 0.460721 I & 29.8935 - 0.250227 I & 0.13932 - 38.8091 I  \\
\hline   
\end{tabular}
}
\label{tableset4}
\end{table*}

\begin{table*} 
\centering
\caption{QNFs for scalar perturbations via the WKB method for $q=0.1$, $\mu=2$, $l=5$ and $k=1.75$ for several values of the non-minimal coupling parameter $\xi$. }
\resizebox{0.8\columnwidth}{!}{%
\begin{tabular}{cccc}
\hline
$n$ &  $\xi=0.1$ & $\xi=0.5$ & $\xi=0.9$ \\
\hline
   0  &  2.17307 - 0.255753 I  &  2.17311 - 0.255756 I & 2.17315 - 0.255758 I  \\
   1  &  2.12878 - 0.772917 I & 2.12882 - 0.772925 I & 2.12887 - 0.772933 I  \\
   2  & 2.04244 - 1.30734 I & 2.04248 - 1.30736 I & 2.04253 - 1.30737 I \\
   3  & 1.91917 - 1.87077 I & 1.9192 - 1.87080 I & 1.91926 - 1.87081 I  \\
   4  & 1.76672 - 2.47407 I & 1.76673 - 2.47415 I & 1.7668 - 2.47414 I  \\
\hline   
\hline
$n$ &  $\xi=1$ & $\xi=10$ & $\xi=100$ \\
\hline
   0   &  2.17317 - 0.255759 I  &  2.17416 - 0.255818 I & 2.18415 - 0.256415 I  \\
    1  & 2.12888 - 0.772934 I & 2.12989 - 0.773112 I & 2.14000 - 0.774892 I  \\
   2  & 2.04255 - 1.30737 I & 2.04358 - 1.30766 I & 2.05395 - 1.31058 I \\
  3  & 1.91928 - 1.87080 I & 1.92035 - 1.8712 I & 1.93113 - 1.87516 I  \\
   4  & 1.76685 - 2.47410 I & 1.76797 - 2.4746 I & 1.77933 - 2.47939 I  \\
\hline   
\end{tabular}
}
\label{tableset5}
\end{table*}

\begin{table*} 
\centering
\caption{QNFs for scalar perturbations via the WKB method for $q=1$, $\mu=2$, $l=5$ and $k=1.75$ for several values of the non-minimal coupling parameter $\xi$. }
\resizebox{0.8\columnwidth}{!}{%

\begin{tabular}{cccc}
\hline
$n$ &  $\xi=0.1$ & $\xi=0.5$ & $\xi=0.9$ \\
\hline
   0  &  2.90347 - 0.325681 I  &  2.90478 - 0.325728 I & 2.90598 - 0.325787 I  \\
   1  &  2.85124 - 0.984032 I & 2.85415 - 0.983627 I & 2.85565 - 0.983695 I  \\
   2  &  2.74539 - 1.66617 I & 2.75648 - 1.66055 I & 2.7596 - 1.65964 I \\
   3  &  2.58331 - 2.39865 I & 2.61962 - 2.36711 I & 2.62801 - 2.36083 I \\
   4  &  2.36511 - 3.22449 I & 2.45482 - 3.10935 I & 2.47515 - 3.08515 I  \\
\hline   
\hline
$n$ &  $\xi=5$ & $\xi=50$ & $\xi=100$ \\
\hline
   0  &  2.91816 - 0.326413 I  &  3.05059 - 0.333757 I & 3.19694 - 0.342039 I  \\
   1  & 2.86902 - 0.985207 I & 3.00056 - 1.00782 I & 3.14697 - 1.03296 I  \\
   2  & 2.7783 - 1.6591 I & 2.90186 - 1.70253 I & 3.04435 - 1.7477 I \\
   3  & 2.66229 - 2.34635 I & 2.75761 - 2.43481 I & 2.88291 - 2.51526 I  \\
   4  & 2.54342 - 3.02458 I & 2.57361 - 3.22566 I & 2.65751 - 3.38674 I  \\
\hline   
\end{tabular}
}
\label{tableset6}
\end{table*}

\end{document}